\documentclass[%
 reprint,
 amsmath,amssymb,
 aps,soul,
prl,
]{revtex4-2}

\usepackage{graphicx}
\usepackage{dcolumn}
\usepackage{bm}
\usepackage[dvipsnames]{xcolor} 

\usepackage{hyperref}
\hypersetup{
    colorlinks=true,        
    linkcolor=MidnightBlue, 
    citecolor=ForestGreen,  
    filecolor=magenta,      
    urlcolor=cyan,          
    breaklinks=true,        
    bookmarksnumbered=true, 
    pdfpagemode=UseOutlines, 
}
\usepackage{xcolor}

\newcommand{\vk}{\bm{k}}
\newcommand{\vx}{\bm{x}}
\newcommand{\vV}{\bm{V}}
\newcommand{\dx}{\Delta\bm{x}}
\newcommand{\Nh}{\widehat{N}}
\newcommand{\E}{\mathbb{E}}

\renewcommand{\selectlanguage}[1]{}
\usepackage{lipsum}
\makeatletter
\newcommand*{\balancecolsandclearpage}{%
  \close@column@grid
  \cleardoublepage
  \twocolumngrid
}
\begin{document}

\preprint{APS/123-QED}

\title{Excursion-set structure factor of the auroral electric field
}

\author{Magnus F Ivarsen}
\email{Contact: magnus.fagernes@gmail.com}
\altaffiliation[Also at ]{Department of Physics, University of Oslo, Oslo, Norway}
\affiliation{Department of Physics and Engineering Physics, University of Saskatchewan, Saskatoon, Canada}%

\author{Kaili Song}
\affiliation{Physics Department, University of New Brunswick, Fredericton, Canada}

\author{Jean-Pierre St-Maurice}
\altaffiliation[Also at ]{Department of Physics and Astronomy, University of Western Ontario, London, Canada}
\affiliation{Department of Physics and Engineering Physics, University of Saskatchewan, Saskatoon, Canada}%

\author{Glenn C Hussey}
\affiliation{Department of Physics and Engineering Physics, University of Saskatchewan, Saskatoon, Canada}%

\begin{abstract}
A spacecraft crossing the aurora cannot tell whether structures in the field-aligned currents are short-lived or merely small. However,  by tracking clusters of coherent radar echoes from aurorae and Fourier-transforming motion-shifted echo pairs across radar snapshots, we can calculate the dynamic structure factor of the point-clouds, thereby quantifying the auroral electric field's spatial power and its persistence from 5 to 40~km, in regions where the radar look-direction is perpendicular to the geomagnetic field. This follows from the fact that coherent E-region radar echoes appear only where the ionospheric electric field exceeds the Farley--Buneman instability threshold, and so a point-target imaging radar naturally samples the field's excursion set; the field's spatiotemporal correlations are discernible from that point process. We find that the echo pattern moves with its cluster: large-scale structures persist for tens of seconds, but scales under 10~km renew within seconds, and conjunctions with Swarm satellite pairs replicate the same scale dependence in field-aligned currents. Coherent radar can thus quantify multi-scale variability in the auroral electric field on Alfv\'enic timescales.
\end{abstract}



\maketitle

\section{Introduction}

The auroral E-layer (90--150 km altitude over Earth's polar regions) hosts some of the most strongly driven plasma turbulence available to direct observation in nature. An externally imposed electric field drives the electrons, but not the ions, into \textit{relative drift}; when that drift exceeds the ion-acoustic speed, the plasma becomes unstable to the Farley--Buneman (FB) instability \cite{farleyPlasmaInstabilityResulting1963,bunemanExcitationFieldAligned1963}. Meter-scale FB waves then form large regions of field-aligned density irregularities spanning meters to tens of kilometers -- the turbulent electrojets -- whose density fluctuations scatter radio waves, rendering the turbulence \textit{visible} as radar aurora \cite{sahr_auroral_1996}. Imaging  radars have begun to resolve this turbulence around aurorae \cite{bahcivanObservationsColocatedOptical2006,hysell_imaging_2008,ivarsen_turbulence_2024}.

The statistical description of auroral turbulence has rested  on  fluctuation spectra. Rockets return one-dimensional spectra of density and electric-field perturbations \cite{mounirSmallscaleTurbulentStructure1991,spicherDirectEvidenceDoubleslope2014}, while coherent radars sample the turbulence directly only at the Bragg scale and infer turbulent behavior through Doppler spectra  \cite{st.-mauriceOriginNarrowNonionacoustic1994,st.-mauriceTheoreticalFrameworkChanging2016,st-mauriceNarrowWidthFarleyBuneman2023}. The spatial spectrum of the irregularity field itself, distinct from the Doppler velocity or the power at the single Bragg wavenumber, has been difficult to obtain \cite{pecseliSpectralPropertiesElectrostatic2015}.

One response is to treat radar scatter as samples of an underlying field and recover that field's second-order statistics, as recently performed in studies of the neutral atmosphere that use meteor-radar Doppler speeds \cite{vierinenObservingMesosphericTurbulence2019,pobletHorizontalCorrelationFunctions2023} and treat echo locations as essentially random.

Rather than being randomly placed, auroral E-region radar echoes originate from plasma that is \textit{Farley--Buneman unstable}. We therefore analyze the \textit{positions} of coherent-scatter echoes, as observed by the 3D radar \textsc{icebear} \cite{huyghebaertICEBEARAlldigitalBistatic2019,huyghebaertPropertiesICEBEARERegion2021,lozinskyICEBEAR3DLowElevation2022}, building on an earlier Monte-Carlo estimate of their two-point statistics \cite{ivarsenDistributionSmallScaleIrregularities2023,ivarsenMeasuringSmallscalePlasma2023,ivarsen_turbulence_2024-1}. The method of correlating point-sets has a long history, from the X-ray scattering of Debye \cite{debyeZerstreuungRontgenstrahlen1915} and Zernike \& Prins \cite{zernikeBeugungRoentgenstrahlenFluessigkeiten1927}, to van Hove's  generalization to Born scattering \cite{vanHoveCorrelationsSpaceTime1954}, the modern classification of point patterns by their density fluctuations \cite{torquatoLocalDensityFluctuations2003}, and the correlated scattering of cloud droplets  \cite{kostinskiSpatialDistributionCloud2000}.

Electrojet echoes form such a point set because they appear only where the electron drift exceeds a value close to the ion-acoustic speed, making radar aurorae natural \textit{excursion sets} of the electric field. From the echo positions, we calculate the dynamic structure factor, i.e., the power spectrum of the echo pattern in space and time, which tells us how much structure exists at each scale, and for how long it persists. We find that echo clusters carry a characteristic pattern that moves with the cluster, that structure at scales of tens of kilometers persists for tens of seconds, and that structure at 5--10~km is renewed within seconds, likely due to the action of dispersive, kinetic Alfv\'en waves \cite{chastonWidthBrightnessAuroral2003,luhr_field-aligned_2015,milesAlfvenicDynamicsFine2018}. The method thereby opens a new route to the structure of auroral turbulence.

\section{Methodology}

Radar aurorae move. Echo regions in the auroral E-region appear to drift at speeds
of order 1~km\,s$^{-1}$ \cite{ivarsen_point-cloud_2024,ivarsen_deriving_2024}. With that example speed, echo positions accumulated over, say, 16~s are smeared over the 16~km displacement that the pattern travels in that time. The \textsc{icebear} 3D radar, however, locates echoes in three dimensions once per
second \cite{lozinskyICEBEAR3DLowElevation2022}, inviting us to treat each 1-s image as a snapshot. We then follow moving echo clusters with an automatic tracker, and account for their translation as a \textit{phase} between snapshots, measured from the data. 

\paragraph{Echoes as an excursion set.}
FB waves grow where the electron drift exceeds the ion-acoustic speed; coherent echoes are therefore seen only where the electric field exceeds a threshold $E_c\approx20$~mV\,m$^{-1}$ \cite{st-mauriceNarrowWidthFarleyBuneman2023,koustovRelationshipVelocityEregion2005}. Writing the field as a perturbation to the mean
convection $E_0$,
\begin{equation}
E = E_0 + \sigma_E F, \label{eq:E}
\end{equation}
with $\sigma_E$ the fluctuation amplitude and $F$ a Gaussian random field of zero mean and unit variance, gives the FB threshold at the dimensionless level, in $F$, of $\nu=(E_c-E_0)/\sigma_E$. The expected echo density is then,
\begin{equation}
\Lambda(\vx,t) = \lambda_0\, g\big(F(\vx,t)\big), \label{eq:Lambda}
\end{equation}
where $\lambda_0$ is a density scale and $g$ is a non-negative function that vanishes for $F<\nu$. The link $g$ describes how the density of saturated meter-scale irregularities grows once the instability threshold
is exceeded.

The irregularities are ephemeral or near-instantaneous compared to the driver \cite{ivarsen_deriving_2024}: individual turbulent regions  die out within a second or two \cite{prikryl_doppler_1988,prikryl_evidence_1990}. Each 1-s radar image is therefore approximately a fresh Poisson draw of echoes with rate $\Lambda$. A Poisson process whose rate is itself a random field is a \textit{Cox process} \cite{coxStatisticalMethodsConnected1955,daleyIntroductionTheoryPoint2006}.

The natural measure of the clustering in $\Lambda$ is the structure factor, the power spectrum of a point pattern. Its space-time generalization, which adds a delay $\tau$ between images (the intermediate scattering function of liquid-state physics \cite{schoenComputationAnalysisDynamic1986}), is \cite{diggleStatisticalAnalysisSpatial2013},
\begin{equation}
S(\vk,\tau)=\delta_{\tau0}+\frac{P_\Lambda(\vk,\tau)}{\Lambda_0},
\label{eq:Skt}
\end{equation}
with $\Lambda_0$ the mean echo density per image and $P_\Lambda(\vk,\tau)$ the spectrum of the covariance of $\Lambda$ between images $\tau$ apart. The term $\delta_{\tau0}$ is shot noise: each echo is perfectly correlated with itself, and this is all that a perfectly structureless pattern would contain. For $\tau\ge1$~s the shot-noise term is absent, because an echo nominally cannot be paired with itself in another image. Spectra formed from pairs of echoes in different images measure the pattern without a noise floor (Appendix~A). Because $g$ acts point by point, a scale expressed in the field appears at the same scale in the echo pattern, whatever $g$ is, while spectral slopes depend on $g$ (Appendix~E).

\begin{figure}
    \centering
    \includegraphics[width=.5\textwidth]{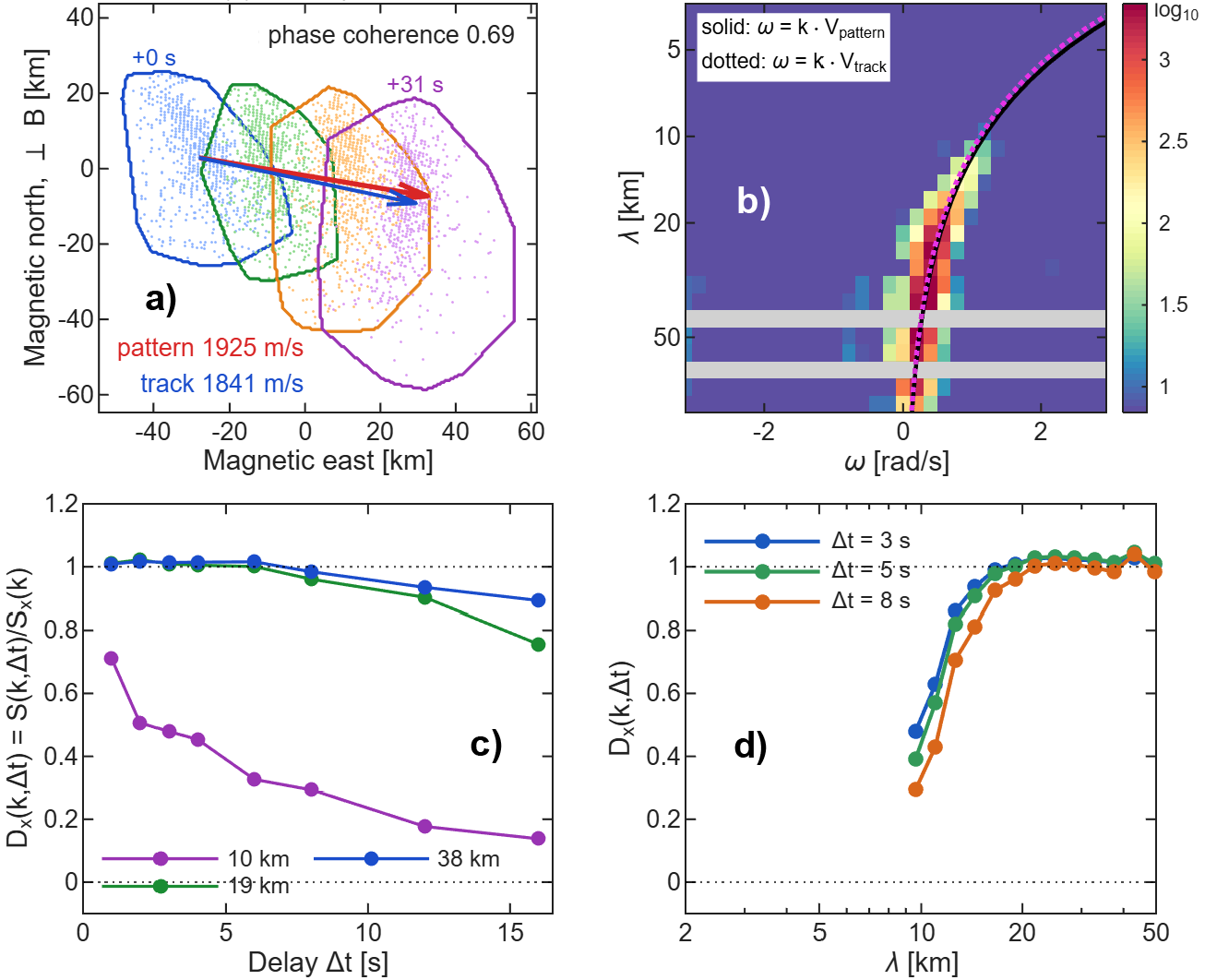}
    \caption{One block of echoes in the frame moving with the echo pattern, observed on 6 May 2023, 11:06:44 UT, for 32~s, with a phase coherence (Eq.~\ref{eq:Cphi}) of 0.69. \textbf{Panel a):} Echoes (dots) and footprints (outlines) of four 1-s images evenly spaced over the block, with the displacements over the block at the pattern velocity (red, 1925~m\,s$^{-1}$) and the tracked velocity (blue, 1841~m\,s$^{-1}$). \textbf{Panel b):} $S(\vk,\omega)-1$ (Eq.~\ref{eq:Skw}). The solid line shows $\omega=\vk\cdot\vV_{\rm pattern}$, the dotted line shows $\omega=\vk\cdot\vV_{\rm track}$, and grey bands denote rings with no modes in this cut. \textbf{Panel c):} Persistence $D(k,\Delta t)$ (Eq.~\ref{eq:D}) against delay at three wavelengths. \textbf{Panel d):} $D$ against wavelength at three delays.}
    \label{fig:ex}
\end{figure}

\paragraph{Lagged spectra in the moving frame.}
The tracker measures the motion of entire echo regions, a proxy for ionospheric electric field enhancements \cite{ivarsen_point-cloud_2024,ivarsen_deriving_2024,ivarsen_eastward_2025-1,ivarsenExtremeTransientBursts2026}. In the present paper, the tracker supplies the \textit{footprint} (as \textsc{matlab} $\alpha$-shapes \cite{edelsbrunnerShapeSetPoints1983}) of an evolving echo cluster, inside of which every echo is subsequently counted; Figure~\ref{fig:ex}a) shows this motion, in four snapshots of a particular block. With $n_s$ the echo counts of image $s$ on a 0.5-km grid, $N_s$ their total, and $w_s$ the footprint's grid cells normalized to unit sum, the term $\delta_s=n_s-N_sw_s$ removes the footprint's own shape. Pairs of echoes within this cluster, in images $\tau$ apart, give the lagged spectrum,
\begin{equation}
S(\vk,\tau)=\frac{\sum_s\mathrm{Re}\big[\hat\delta_s(\vk)\,\hat\delta^*_{s+\tau}(\vk)\,e^{-i\tau\vk\cdot\vV}\big]}{\sum_s\big(\Nh_s\Nh_{s+\tau}\big)^{1/2}},
\label{eq:Slag}
\end{equation}
where,
\begin{equation}
\hat\delta_s(\vk)=\sum_{\vx}\delta_s(\vx)\,e^{-i\vk\cdot\vx},
\label{eq:debye}
\end{equation}
is the point-set form of the Debye scattering relation
\cite{debyeZerstreuungRontgenstrahlen1915,bartlettSpectralAnalysisTwodimensional1964,hawatEstimatingStructureFactor2023}. In Eq.~(\ref{eq:Slag}), $\Nh_s$ is the count within the cluster that image $s$ would hold if echoes were spread uniformly over the cluster's footprint at mean echo density and allowed to drift linearly in time, so as to keep the correlation of the counts themselves out of the lagged spectrum. The summation index $s$ runs over a block of 32 or 64 consecutive images. 

After a time $\tau$, a pattern translating at a speed $\vV$ would shift the image by the distance $\vV\tau$, which gives the product $\hat\delta_s\hat\delta^*_{s+\tau}$ a phase $e^{+i\tau\vk\cdot\vV}$. The factor $e^{-i\tau\vk\cdot\vV}$ in Eq.~(\ref{eq:Slag}) subsequently removes the phase accumulated by this product. Translation alone therefore leaves $S(\vk,\tau)$ unchanged, and, importantly, any decay with $\tau$ is a \textit{change of the structure itself}, in the frame moving with the pattern. From this, three quantities follow.

\paragraph{(1) Pattern velocity.}
Multiplying each product $\hat\delta_s(\vk)\,\hat\delta^*_{s+\tau}(\vk)$ by the phase $e^{-i\tau\vk\cdot\vV'}$ of a trial velocity $\vV'$ will only give real and positive summed products when the trial velocity $\vV'$ is the velocity of the pattern itself. For any other trial velocity, the remaining phases will result in a smaller real sum. The pattern velocity is therefore the trial velocity for which their sum is largest,
\begin{equation}
\vV=\arg\max_{\vV'}\;\sum_{\tau=1}^{8}\sum_{\vk}\mathrm{Re}\big[\sum_s\hat\delta_s\hat\delta^*_{s+\tau}\,e^{-i\tau\vk\cdot\vV'}\big],
\label{eq:V}
\end{equation}
which we evaluate over wavelengths from 5 to 40~km. The phase coherence $C_\phi$ (Appendix~B) is 1 when every mode moves at $\vV$ and near 0 when the phases are unrelated.

The pattern velocity is the motion of the structure within an echo cluster, measured from the phases of the echo positions \textit{inside} the footprint; the tracked velocity is the motion of the cluster's centroid. The search for the pattern velocity uses the tracked motion as a first trial, but extends
$\pm1$~km\,s$^{-1}$ around it (Appendix~B), and so agreement is never imposed. Comparing the two velocities shows whether a cluster's fine structure is carried \textit{with} the cluster, which we expect if both are moved by the same $\bm E\times\bm B$ velocity. The motion is  visible directly in the frequency spectrum of the image sequence, $S(\vk,\omega)$, where a translating pattern puts its power on the ridge $\omega=\vk\cdot\vV$ (Appendix~B). Figure~\ref{fig:ex}b) shows an example where the ridge in $S(\vk,\omega)$ and the expectations from the tracked velocity agree within 5\% (see also Figure~\ref{fig:result1}).

\paragraph{(2) Equal-time spectrum.}
Pairs of echoes within one image also see structure that is gone a second later, which pairs in different images do not (Appendix~C). We therefore take the equal-time spectrum of the structure that persists by extrapolating the lagged spectra back to zero delay,
\begin{equation}
S(\vk)=\frac{S(\vk,1~\mathrm{s})^2}{S(\vk,2~\mathrm{s})},
\label{eq:Sx}
\end{equation}
which is log-linear in $\tau$, and so gives the equal-time spectrum smoothed over the three images (3 s) that the two lags span, without the shot-noise term of Eq.~(\ref{eq:Skt}). Each spectrum of Eq.~(\ref{eq:Sx}) is corrected for the radar's resolution using an instrument transfer function (Appendix~A) and averaged over rings of constant $k=|\vk|$. We show the reduced spectrum $k\,S(k)$, since $S(k)$ is an areal spectrum. This allows for direct comparison with \textit{in-situ} measurements; a satellite measures along its orbital path, and the spectrum along a straight cut through an isotropic two-dimensional field has the slope of $k\,S(k)$ rather than that of $S(k)$ (Appendix~D).

\paragraph{(3) Persistence.}
The lagged spectra also contain information about how long structure of each size \textit{persists}. The fraction of the structure at wavenumber $k$ that is still present after a delay $\Delta t=\tau$, in the frame of the pattern, is,
\begin{equation}
D(k,\Delta t)=\frac{S(k,\Delta t)}{S(k)},
\label{eq:D}
\end{equation}
with $S(k)$ the equal-time spectrum of Eq.~(\ref{eq:Sx}). $D=1$ means that nothing at that scale has changed, and $D=0$ that the pattern is unrelated to its earlier self. The persistence $D$ is that of the echo pattern. At a single point the echo correlation decays at least as fast as the field's (Appendix~E). In Appendix~C, we describe in detail how $D$ is formed; Figure~\ref{fig:ex}c) shows $D(k,\Delta t)$ for the evolving cluster, demonstrating a rapid loss in 10-km structure, while structure on scale-sizes $\gtrsim20$~km clearly persists; Figure~\ref{fig:ex}d) shows that a scale-size around 15~km separates the two regimes.

\paragraph{Satellite conjunctions.}
During conjunctions with the European Space Agency's Swarm satellites \cite{friis-christensenSwarmConstellationStudy2006} we compute the spectrum of the magnetic field residual from the vector field magnetometer \cite{merayoSwarmMagnetometryPackage2008} (Appendix~D). Swarm A and C largely orbit side by side and cross the same current sheets seconds apart, and the along-track derivative of the transverse magnetic perturbation is proportional to the field-aligned current (FAC) density (Amp\`ere's law for a current sheet \cite{rotherCHAMPObservationIntense2007,ritterDeterminingFieldalignedCurrents2013}). A scale-dependent coherence measurement then reflects the fraction of FAC structure that survives the delay between the two satellites \cite{luhr_field-aligned_2015,yangStatisticalCorrelationAnalysis2018}; it involves no threshold. We compare satellite-measured coherence with the radar persistence $D$ (Appendix~D).

\section{Results}

\begin{figure}
    \centering
    \includegraphics[width=.5\textwidth]{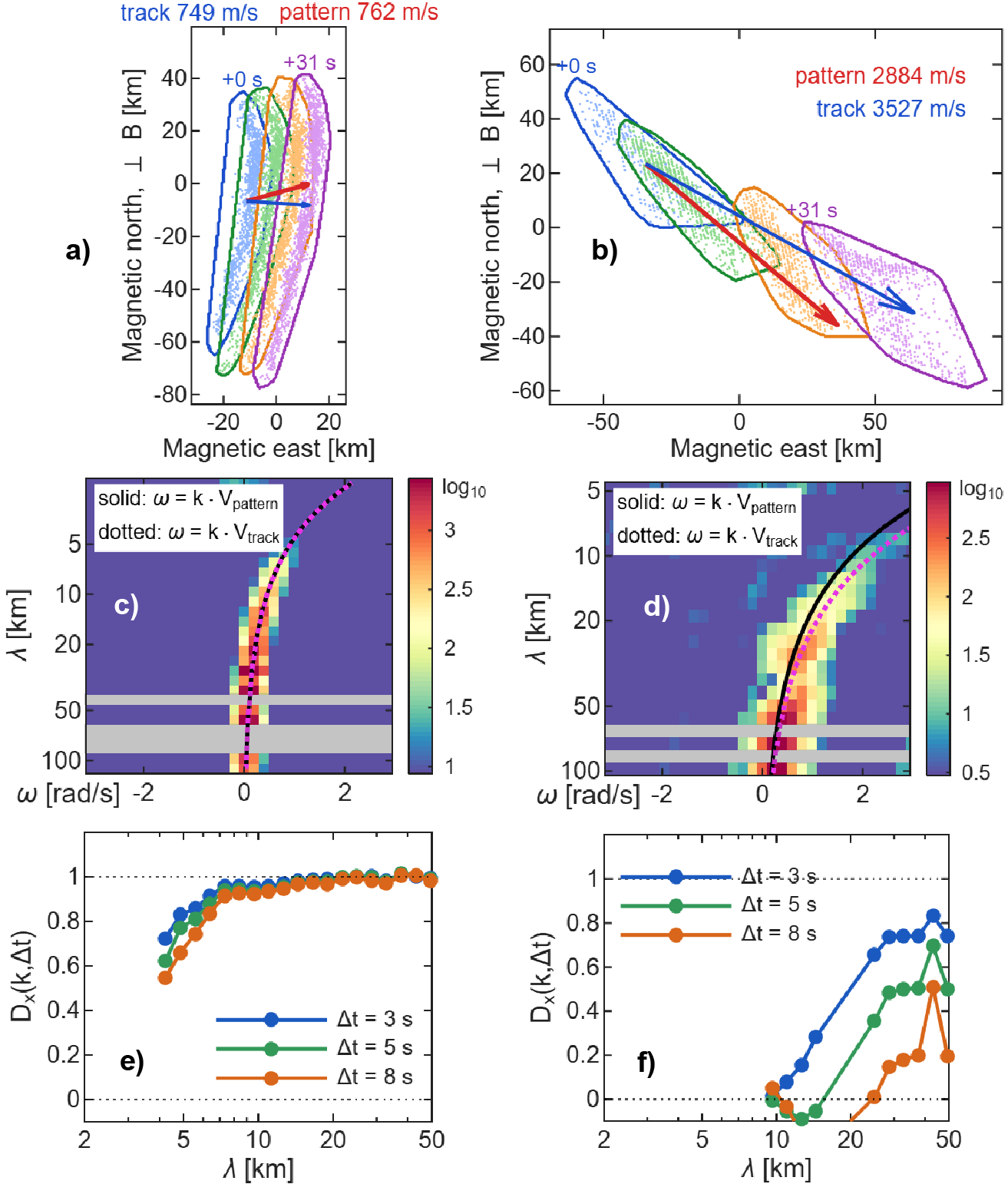}
    \caption{Two clusters, one persistent (left, 7 Jul 2023, 08:42:08~UT, $C_\phi= 0.99$), and one rapidly renewing (right, 13 Mar 2021, 05:18:51~UT, $C_\phi= 0.56$). \textbf{Panels a, b):}  Echoes and footprints of four 1-s images, with the pattern (red) and tracked (blue) displacements over the 32-s block.  \textbf{Panels c, d):} $S(\vk,\omega)-1$ for wavevectors along the pattern velocity, with $\omega=\vk\cdot\vV_{\rm pattern}$ (solid line) and $\omega=\vk\cdot\vV_{\rm track}$ (dotted line).  \textbf{Panels e, f):} $D(k,\Delta t)$ against wavelength for $\Delta t=3$, 5 and 8~s.  
    }
    \label{fig:result1}
\end{figure}

\begin{figure}
    \centering
    \includegraphics[width=.5\textwidth]{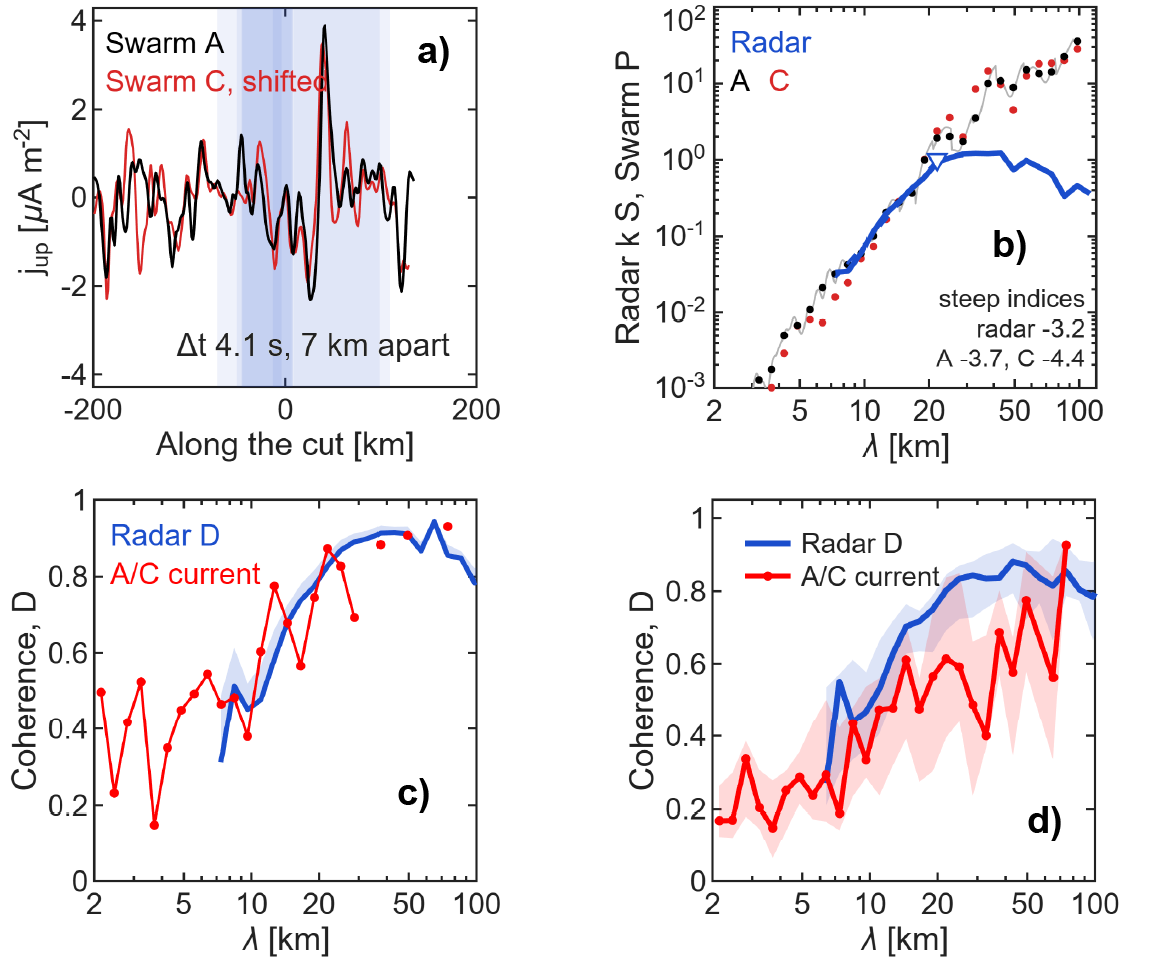}
    \caption{Swarm A--C over \textsc{icebear} on 28 August 2021, at 07:04 UT. \textbf{Panel a):} Upward field-aligned current along the track from Swarm A (black) and C (red), with C shifted by 4.1~s (correlation 0.96; Appendix~D).  Shading: the mapped latitudes of the radar structures.  \textbf{Panel b):} Radar structure factor at zero delay, $k\,S(k)$ (blue), and the Swarm spectra of the magnetic residual normal to the cut from A (black dots and line) and C (red dots). \textbf{Panel c):} Coherence of the A--C currents per ring (black dots and line), and the fraction of the radar structure kept over the same 4.1~s in the frame moving with the pattern (blue).  \textbf{Panel d):} The same as panel c), for all 11 conjunctions identified; shaded regions denote quartiles (see Figures S1--S3 in the Supplemental Material). 
    }
    \label{fig:result2}
\end{figure}

Figure~\ref{fig:result1} contrasts two highly dissimilar clusters, one moving eastward at 749~m\,s$^{-1}$, with a pattern speed (Eq.~\ref{eq:V}) of 762~m\,s$^{-1}$, and one almost continuously \textit{crashing} in the south-east direction at 3527~m\,s$^{-1}$, with a pattern speed of 2884~m\,s$^{-1}$. The former shows persistence $D>0.9$ for scales down to around 7~km, after which structure starts to renew; the latter exhibits rapid renewal of structure for all scales, except for the largest scales ($\gtrsim40$~km) and shortest lags (3~s). Both show a cliff in $D$ as a function of scale.

Space-ground conjunctions elucidate the mechanisms at play. Figure~\ref{fig:result2}a) shows the FAC density from Swarm A during one crossing of \textsc{icebear}'s field of view, and from Swarm C shifted 4.1~s back (correlation 0.96). The spectrum of the magnetic residual (Figure~\ref{fig:result2}b) matches the radar $k\,S(k)$ below about 20~km, both very steep (dissipative). At the same 4.1-s delay, $D$ follows the A--C coherence from about 7 to 50~km (Figure~\ref{fig:result2}c); of the 11 Swarm A--C conjunctions with \textsc{icebear} that we have identified, this one had the smallest separation between A and C (6.6~km). Both instruments tend to see rapid renewal of $\lesssim10$~km structure  (Figure~\ref{fig:result2}d).

\begin{figure}
    \centering
    \includegraphics[width=.5\textwidth]{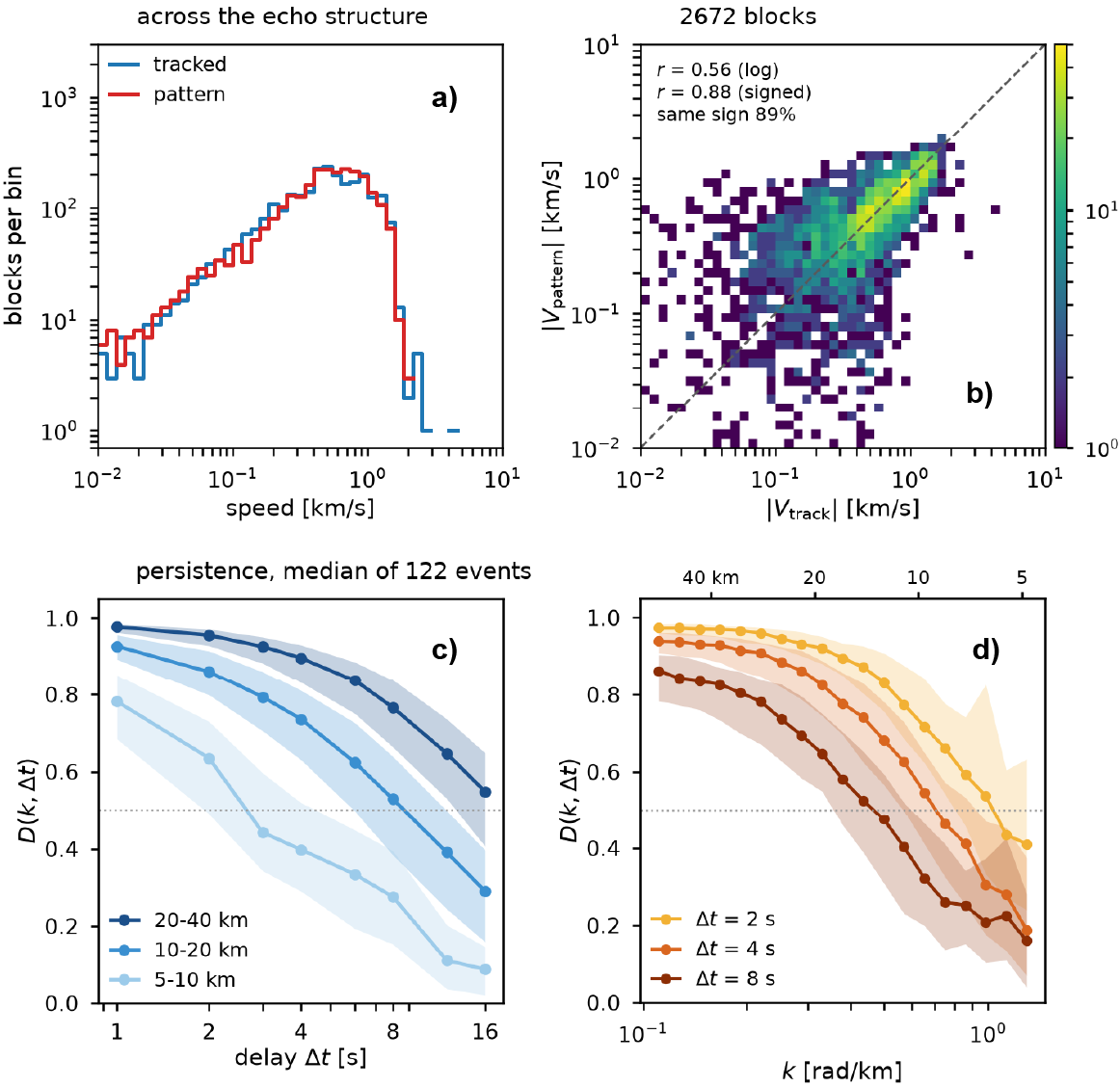}
    \caption{
    \textbf{Panel a):} Distributions of tracked (black) and pattern (red) speed, for the direction transverse to the cluster's elongation. \textbf{Panel b):} Pattern against tracked speed for $2672$ blocks with phase coherence above 0.35, with a 1:1 relation indicated (dashed). \textbf{Panel c):}  Persistence $D(k,\Delta t)$ as a function of  delay $\Delta t$ for three scale-intervals: dotted lines indicate median over blocks, with quartiles shown with a shaded region. \textbf{Panel d):} Persistence $D(k,\Delta t)$ at $\Delta t=3$, 5 and 8~s; as a function of wavenumber (bottom) and scale-size (top). See Video S1 in the Supplemental Material for an event video.
    }
    \label{fig:stats}
\end{figure}

For Figure~\ref{fig:stats}, we searched a database of 903 radar days in the years 2020–2025 for events in which most echoes belong to long-lasting tracks, defining an event as the interval where at least 60\% of the echoes are successfully tracked, at a rate of at least 1000 echoes/s, for more than 120~s. This gives  129 events, falling on 52~days; their median duration is 205~s, their median rate 1400~echoes/s, and their median share of tracked echoes is 85\%. The 129 events featured a median auroral electrojet index value  \cite{gjerloevSuperMAGDataProcessing2012} of 900~nT, meaning they represent extremely disturbed conditions in geospace (98$^\text{th}$ percentile overall). For each of the 2672 reasonably coherent ($C_\phi>0.35)$ blocks, we compute the elongation of the block's echo cloud: the variance of the echo positions along axes rotated through all azimuths about the cloud’s centre-of-mass, taking the direction of maximum variance as the echo axis. A cloud whose ratio of largest to smallest variance exceeds 2 is elongated; wholly 93\% of the blocks are elongated (median ratio 6.8), and on scales of 20–45 km, spectral power lies at wavevectors transverse to the axis. In the transverse direction, the tracked and pattern speeds agree: their distributions are nearly identical (medians 0.46 and 0.48 km/s; Figure~\ref{fig:stats}a), and across the 2672 blocks they correlate with $r = 0.88$ (signed components) and $r = 0.56$ (logarithms of the magnitudes), with the same sign in 89\% (Figure~\ref{fig:stats}b). For the direction along the echo elongation, the pattern velocity tends to be faster than the tracked motion.

Lastly, Figure~\ref{fig:stats}c, d) show median curves for persistence $D$ per delay and scale-size for the intervals, showing an unequivocal collapse in $D$ below a scale-size around 10--15~km. At 5~s, the median $D$ of about 0.4 at 5--10~km gives an $e$-folding time of about 5~s, well above the 1-s exposure of an image.

\section{Discussion}

\label{sec:discussion}

Coherent radar echoes from the E-region require FB threshold exceedance. A point-target imaging radar therefore samples the \textit{excursion set} of the electric field. Because the threshold sampling preserves the scales and the motion of the field that drives echo occurrence (Appendix E), the radar can quantify, scale by scale, the structure of the auroral electric field and how it moves. The persistence of the echo pattern is tied to the underlying field by the satellite comparison (Fig.~\ref{fig:result2}). While satellites can separate such temporal evolution from spatial variation only statistically \cite{matthaeusEnsembleSpaceTimeCorrelation2016}, the point-target imaging radar now observes it directly.

Some limitations qualify the reading. The $|\vk|$-ring average invokes statistical isotropy in the thin-layer plane perpendicular to $\bm B$, whereas the irregularities are anisotropic, aspect-sensitive and field-aligned \cite{kudekiAspectSensitivityEquatorial1989} and the echo clusters themselves are elongated; aspect sensitivity acts as an envelope on the echo rate, which the expected count removes to first order, but a cluster drifting through the radar's field of view changes visibility in time, and echoes come only from directions perpendicular to the geomagnetic field.

However, the sharp drop in echo persistence below $\sim10$~km appears also in the Swarm A--C coherence of the FAC (Fig.~\ref{fig:result2}c, d) and in earlier Swarm studies; we therefore attribute it to the field itself. Ref.~\cite{luhr_field-aligned_2015} cross-correlated FAC densities and found that large-scale structures tend to persist, whereas smaller-scale ones do not. Later studies have similarly found that inter-spacecraft correlation drops significantly for small scale-sizes compared to larger ones \cite{forsythIdentifyingIntervalsTemporally2017,yangStatisticalCorrelationAnalysis2018,zhouPropertiesLargeamplitudeKilometerscale2025,luhr_small-scale_2025}. What is more, kilometer-size FACs exhibit very rapid ($<1$~s) spikes in current amplitude \cite{luhr_small-scale_2025}. We point to kinetic (dispersive) Alfv\'{e}n waves interacting with the auroral ionosphere \cite{chenKineticAlfvenWave2013,dimareNewRegimeInertial2024}. Arcs tend to exhibit characteristic scale-sizes of 2--20~km \cite{knudsenWidthStructureMesoscale2001}, seen in intense kilometer-scale FACs \cite{rotherCHAMPObservationIntense2007,zhouPropertiesLargeamplitudeKilometerscale2025}, and also in auroral radar studies \cite{kelleyOriginSpatialExtent1982,rinoRecentStudiesStructure1983,streltsovCouplingDensityStructures2008,st.-mauriceTheoreticalFrameworkChanging2016,forsytheAsymmetryFarleyBunemanDispersion2016,ivarsenAlgorithmSeparateIonospheric2023,huyghebaert_investigating_2023}. Km-size arc-adjacent structure carries Alfv\'enic perturbations of the same size \cite{milesAlfvenicDynamicsFine2018,greene_situ_2025}, subject to rapid reconfiguration by kinetic magnetohydrodynamic turbulence \cite{wygantEvidenceKineticAlfven2002} and its attendant particle precipitation \cite{lysakKineticAlfvenWaves2023a}.  

\section{Conclusion}

Coherent E-region radar echoes form excursion sets: they mark where the electric field exceeds the Farley--Buneman threshold. The observed echo rate therefore forms a \textit{Cox process} \cite{coxStatisticalMethodsConnected1955}, whose dynamic structure factor we have measured directly using radar imaging. We find that echo clusters carry a characteristic pattern that moves with the cluster at around the tracked velocity (Fig.~\ref{fig:stats}), at times as a single coherent object and at times like a crashing wave (Fig.~\ref{fig:result1}). Echo patterns persist for tens of seconds at scale-sizes that measure tens of kilometers, while at 5--10~km, structure is \textit{continuously renewed}. Pairs of Swarm satellites see the same loss of small-scale structure in the field-aligned current; we therefore deduce that the renewal is a property of the auroral fields. 

A natural next step is the synthetic recovery of $D$ with a known decay law, a step towards structure lifetimes \cite{hysellDecayingEquatorialRegion1997,ivarsenLifetimesPlasmaStructures2021,ivarsen_plasma_2024}. 
At present, the results have direct applications: single-spacecraft estimates of field-aligned currents assume that a structure does not change the currents are being traversed \cite{forsythIdentifyingIntervalsTemporally2017}. Coherent radar measures, scale by scale, how long that assumption holds. The scale-dependent persistence can constrain models of dispersive Alfv\'en waves. Coherent E-region radars can thereby directly estimate how the auroral electric field evolves in time, scale by scale.

\vspace{-5pt}
\section*{Acknowledgements}
\vspace{-17pt}
This work is supported by the European Space Agency (ESA) grant no. 1000012348. We also acknowledge the support of the Canadian Space Agency (CSA) [20SUGOICEB], the Canada Foundation for Innovation (CFI) John R. Evans Leaders Fund [32117], the Natural Sciences and Engineering Research Council of Canada (NSERC), the Discovery grants program [RGPIN-2025-04351], the Digital Research Alliance of Canada [RRG-4802]. MFI is grateful to J Vierinen, O Nestande, and I Ivarsen for stimulating discussions. See Supplemental Material [URL] (containing Ref.~\cite{knudsenThermalIonImagers2017}) for detailed conjunction descriptions and an event video. Anthropic's Claude Fable 5, Fable 5.1, Opus 5.5, and Google Scholar Labs were used to assist research and coding in \textsc{matlab}, at the sole responsibility of the authors.

\vspace{-14pt}
\section*{Data availability}
\vspace{-17pt}
\textsc{icebear} 3D echo data for 2020, 2021 is published with DOI \url{10.5281/zenodo.7509022}. Data from the European Space Agency's Swarm mission is available at \url{https://swarm-diss.eo.esa.int}. \textsc{matlab} code that performs the present analysis is published with DOI \url{10.5281/zenodo.20810817}.

\balancecolsandclearpage


\appendix

\section{End Matter}

\subsection{Appendix A: data and estimator}

We use the auroral coherent scatter radar \textsc{icebear}'s \cite{huyghebaertICEBEARAlldigitalBistatic2019} 3D echo product, where each echo comes from one range-azimuth-Doppler bin of a 1-s integration, which \textsc{icebear} treats as an image (about 200,000 per second) whose elevation the imaging determines; the echoes of one second form what we call a 1-s image. Echo locations are imaged by multiple-frequency interferometry \cite{lozinskyICEBEAR3DLowElevation2022}. Echoes between 90 and 130~km altitude and above 1~dB signal-to-noise ratio are carried along the field line to 105~km, onto the plane perpendicular to the magnetic field. The field direction comes from the echoes themselves: a least-squares fit of their horizontal positions against altitude and field-line-traced (AACGM \cite{bakerNewMagneticCoordinate1989}) offsets gives the horizontal displacement per kilometer along the field line. All distances are measured in this plane.

The tracker \cite{ivarsen_point-cloud_2024} clusters the echoes of each 1-s frame within a 3-s window with an adaptive DBSCAN, represents each cluster by its $\alpha$-shape \cite{edelsbrunnerThreedimensionalAlphaShapes1994}, assigns clusters between frames using kinematic prediction \cite{bewleySimpleOnlineRealtime2016,kuhnHungarianMethodAssignment1955}, and fits each track's centroid trajectory with piecewise-linear segments \cite{keoghOnlineAlgorithmSegmenting2001}. A tracked velocity component below 0.3~km\,s$^{-1}$ is set to zero. Each block holds 32 or 64 consecutive images of one track. The footprint of image $s$ is the $\alpha$-shape of the tracker's 3-s frame, with its members carried to the image time along the tracked drift; only the window is built this way, and the counts $n_s$ include every echo inside the cluster from image $s$.

Subtracting the uniformly filled footprint removes the window exactly: for structureless echoes $\E|\hat\delta_s|^2=N_s(1-|\hat w_s|^2)$, and products of different images vanish on average, so that the footprint's sharp edge does not enter the lagged spectra. Pairs of echoes from different images contain no self-pairs, and, given $\Lambda$, the echoes of one image are drawn independently of those of another, so their Poisson scatter is uncorrelated: for $\tau\ge1$~s, the point process adds noise to the estimate but no floor.

Each spectrum is divided, mode by mode, by the radar's transfer function \cite{bracewellFourierTransformIts1986},
\begin{equation}
T(\vk)=\mathrm{sinc}^2\frac{k_r\delta r}{2}\;\mathrm{sinc}^2\frac{k_a\delta a}{2}\;
\mathrm{sinc}^2\frac{k_x\delta r}{2} \times\mathrm{sinc}^2\frac{k_y\delta r}{2}, 
\label{eq:T2D}
\end{equation}
with $\mathrm{sinc}\,u=\sin u/u$, range gate $\delta r=0.75$~km, and  azimuthal resolution $\delta a=R\times0.1^\circ$ at range $R$. 

\subsection{Appendix B: motion}

Equation~(\ref{eq:V}) is maximized on a grid of trial velocities within $\pm1$~km\,s$^{-1}$ of the tracked drift, recentered while its maximum lies on the edge, then refined by Newton steps. The lags run from 1~s to 8~s. The wavelengths run from 5~km, or the aliasing limit $2(|\vV_{\rm track}|+0.5~\mathrm{km\,s^{-1}})\times1$~s where that is larger, to the smaller of 40~km and a third of the footprint. The phase coherence is the magnitude of the sum in Eq.~(\ref{eq:V}) at
$\vV$ over the sum of the magnitudes of its terms,
\begin{equation}
C_\phi=\frac{\big|\sum_{\tau,\vk}\sum_s\hat\delta_s\hat\delta^*_{s+\tau}\,e^{-i\tau\vk\cdot\vV}\big|}{\sum_{\tau,\vk}\big|\sum_s\hat\delta_s\hat\delta^*_{s+\tau}\big|}.
\label{eq:Cphi}
\end{equation}

Transforming the image sequence in time, without removing the motion, gives the dynamic structure factor in frequency,
\begin{equation}
S(\vk,\omega)-1=\frac{\big|\sum_sh_s\,\hat\delta_s(\vk)\,e^{i\omega s}\big|^2}{\sum_s h_s^2N_s}-1,
\label{eq:Skw}
\end{equation}
with $h_s=\frac12[1-\cos(2\pi s/(M-1))]$ the Hann weights over a block of $M$ images, to avoid spectral leakage. The subtracted $1$ is, approximately, the shot noise. Along $\vV$ the ridge $\omega=\vk\cdot\vV$ rises with slope $|\vV|$, across $\vV$ it lies at $\omega=0$, and change of the structure in the moving frame broadens it in proportion to the rate of change. Equation~(\ref{eq:V}) is the lagged form of the power of $S(\vk,\omega)$ on the line $\omega=\vk\cdot\vV'$ summed over wavenumbers. We then test whether the ridge lies on the line $\omega=\vk\cdot\vV$ at \textit{each wavenumber separately}: the ridge located ring ($|\vk|$) by ring, without removing the motion, gives the same velocity at all scales, which would entail that one velocity fits every wavenumber (rigid translation of the pattern). By contrast, structure that is gone within a second, including the excess within single images (Appendix~C), spreads over all frequencies: it raises the floor and leaves the ridge in place.

\subsection{Appendix C: the spectrum and persistence}

Equation~(\ref{eq:Sx}) avoids the structure that is gone within a second (fast-renewing field structure, or instrumental noise) by using pairs from different images only. A ring enters Eq.~(\ref{eq:Sx}) only where $S(\vk,1~\mathrm{s})$ and $S(\vk,2~\mathrm{s})$ both exceed three bootstrap errors and their ratio lies between 0.3 and 1.2, a criterion that truncates the spectra near 5~km.

For Fig.~\ref{fig:stats}c, d) and the conjunctions, persistence, Eq.~(\ref{eq:D}), is formed on the lagged spectra summed over the blocks of an active interval (for conjunctions, the blocks nearest in time to the pass) and weighted by echo counts, ring ($|\vk|$) by ring; intermediate delays are interpolated log-linearly, and the spread we show is that between intervals. Its reference $S(k)$ is the extrapolation of Eq.~(\ref{eq:Sx}), smoothed over 3~s; ratios of $D$ at two delays do not depend on it.

A single velocity per block moves all wavevectors, so velocity error and shear lower $D$, most at small scales and long delays. Errors come from a bootstrap over 8-s sub-blocks, which includes the extra variance of correlated pairs \cite{hawatEstimatingStructureFactor2023} but treats the sub-blocks as independent, and so understates the error at scales that persist longer than 8~s. Lastly, marginal growth near the edges of a cluster makes the echo rate there change faster than the field itself. The persistence $D$ is therefore strictly that of the echo pattern; Appendix~E shows what carries over to the field exactly, as well as the satellite comparison.

\subsection{Appendix D: Swarm}
Following Refs.~\cite{parkAlfvenWavesAuroral2017,ivarsenObservationalEvidenceRole2020}, we subtract a 3.75-minute second-order Savitzky--Golay filter from the 50-Hz magnetic data of Swarm (polar orbits at 450--500~km), and project the residual onto the plane perpendicular to the geomagnetic field. We then compute the spectrum of its component normal to the satellite's path over 90~s \cite{trobsImprovedSpectrumEstimation2006}. Frequency $f$ then becomes wavenumber, $k=2\pi f/v$, through the speed $v$ of the magnetic footpoint at 105~km relative to the moving pattern (see, e.g., Ref.~\cite{fredricksAmbiguitiesDeductionRest1976}). Using the speed relative to the pattern, not to Earth, places the satellite's samples at the spatial scales of the structure the radar sees. For an isotropic field with $S\propto k^{-n}$, the spectrum along a straight cut is $\int S\,\mathrm{d}k_\perp\propto k^{-(n-1)}$, the slope of $k\,S(k)$. For the Swarm A--C pair,  we shift C by the delay $\Delta t$ at which its residual best matches A's over 10 to 200~km. We then take first differences of both series (the current profile along the path), and form the coherence per ring,
\begin{equation}
\gamma(k)=\frac{\big|\langle X_A X_C^*\rangle\big|}{\langle|X_A|^2\rangle^{1/2}\,\langle|X_C|^2\rangle^{1/2}},
\label{eq:gamma}
\end{equation}
where $X_A$ and $X_C$ are Fourier transforms of the two profiles in 20-s windows and $\langle\cdot\rangle$ averages over windows and over the frequencies in the ring \cite{welchUseFastFourier1967}. Coherence is an amplitude ratio, 1 for profiles that are the same at that scale and $\sim(n_{\rm seg}n_{\rm bins})^{-1/2}$ for unrelated ones, with $n_{\rm seg}$ windows and $n_{\rm bins}$ frequencies in the ring; we draw that level beside the data. Its counterpart in the radar is $D(k,\Delta t)$ of Eq.~(\ref{eq:D}), likewise an amplitude ratio of a cross-spectrum to the equal-time spectrum, taken at the Swarm delay $\Delta t$ from the radar blocks nearest in time and space to the pass. In our case studies, the radar and satellite data are displaced by some minutes and tens of kilometers. The comparison is therefore between the state of the same region, not between simultaneous measurements of the same point. The A--C coherence also falls with the footpoints' separation, so it understates the persistence of the current.

\subsection{Appendix E: what survives an unknown link}

Let $F$ be Gaussian (which is reasonable, as any field with a continuous one-point distribution is a monotone function of a Gaussian field), with correlation $\rho(\dx,\tau)$ between points $\dx$ and $\tau$ apart. For any link $g$ that acts point-by-point, that is square-integrable under the Gaussian measure, then Mehler's expansion \cite{mehlerUeberEntwicklungFunction1866} will give the pair correlation $h$ of the echoes, defined as in Ref.~\cite{kostinskiSpatialDistributionCloud2000} by the factor $1+h$ that enhances the probability of a second echo, as
\begin{equation}
h(\dx,\tau)=\frac{\E[g(F_1)\,g(F_2)]}{\E[g(F)]^2}-1=\sum_{n\ge1}A_n\,\rho^n,
\quad A_n\ge0,
\label{eq:mehler}
\end{equation}
where $A_n=c_n^2/(n!\,c_0^2)$ and $c_n=\E[g(F)\,He_n(F)]$. For a known $g$, as in digital correlators, such as those used in incoherent scatter radar science, this is the arcsine law and its generalizations \cite{hagenDigitalcorrelationTechniquesRadio1973}, inverted to recover $\rho$. Here, the point-wise link fixes the function $h=\mathcal H(\rho)$, the same at every separation and delay, increasing for $\rho\ge0$; the field then fixes $\rho$. Wherever $\rho\ge0$, equal values of $h$ therefore mark equal values of $\rho$: contours of the echo correlation are contours of the field's, and a break at a given scale in the field appears at that scale in the echo pattern. The local slope of $h$, however, is $\rho\,\mathcal H'/\mathcal H$ times that of $\rho$. As for motion, if the field translates, so that $\rho(\dx,\tau)=\rho_m(\dx-\vV\tau,\tau)$, then,
\begin{equation}
h(\dx,\tau)=\mathcal H\big(\rho_m(\dx-\vV\tau,\tau)\big).
\end{equation}
The echo pattern moves with the field's velocity whatever $g$ is. At zero separation in the moving frame, we have persistence, 
\begin{equation}
\frac{h(0,\tau)}{h(0,0)}=\sum_nA_n\rho^n/\sum_nA_n\le\rho(\tau),
\end{equation}
for $0\le\rho\le1$, since $\rho^n\le\rho$: the echo pattern decorrelates at least as fast as the field, and the faster it does so, the more its structure rests on high powers of $\rho$. Scale by scale, the bound is guaranteed only where the field's decay rate grows no faster than $k$: the $n$th term of Eq.~(\ref{eq:mehler}) at wavenumber $\vk$ is a product of $n$ field modes whose wavevectors sum to $\vk$, so its decay is the product of theirs. Otherwise, harmonics of long-lived larger structure can outlive the field's own small-scale structure. 

Whether the echo density appears abruptly at the threshold [an indicator link $g=\mathbb I(F>\nu)$] or grows continuously from zero with the excess above
it [a ramp link $g=(F-\nu)_+$] is a physical question. Several lines of evidence favor the continuous ramp link. FB growth vanishes at marginal stability \cite{fejer_ionospheric_1980}. Waves are driven back to the threshold speed by the nonlocal increase of their aspect angles \cite{drexlerNewInsightsNonlocal2002,st.-mauriceNewNonlinearApproach2001}, and electron heating and echo occurrence both rise continuously above threshold
\cite{st-mauriceRevisitingBehaviorERegion2021,prikryl_doppler_1988,prikryl_evidence_1990,ivarsen_characteristic_2025}. For the ramp, the smoothness of $\rho$ survives in $h$ \cite{PriceUsefulTheoremNonlinear}; for the indicator it does not \cite{vanvleckSpectrumClippedNoise1966}. The invariants that we describe above hold for either link.

\balancecolsandclearpage



%

\end{document}